# Introducing Aspect-Oriented Programming in Improving the Modularity of Middleware for Internet of Things

Senthil Velan S
*Department of Computer Science and Engineering*
*Amity University*
Dubai, UAE
svsugana@gmail.com

*Abstract*—Internet of Things (*IoT*) has become the buzzword for the development of *Smart City* and its applications. In this context, development of supporting software forms the core part of the *IoT* infrastructure. A Middleware sits in between the IoT devices and interacts between them to exchange data among the components of the automated architecture. The Middleware services include hand shaking, data transfer and security among its core set of functionalities. It also includes cross-cutting functional services such as authentication, logging and caching. A software that can run these Middleware services requires a careful choice of a good software modelling technique. Aspect-Oriented Programming (*AOP*) is a software development methodology that can be used to independently encapsulate the core and cross-cutting functionalities of the Middleware services of the *IoT* infrastructure. In this paper, an attempt has been made using a simulation environment to independently model the two orthogonal functionalities of the Middleware with the focus to improve its modularity. Further, a quantitative measurement of the core design property of cohesion has been done to infer on the improvement in the reusability of the modules encapsulated in the Middleware of *IoT*. Based on the measurement, it was found that the modularity and reusability of functionalities in the Middleware software has improved in the *AspectJ* version compared to its equivalent *Java* version.

## I. INTRODUCTION

In today's highly connected world of sensory and interconnected components, efficient communication plays a major role in improving the quality of networked devices, operating in tandem with each other. Wireless network communication between the disparate components have become popular today, because of the growth in high speed communication equipments and related technologies. Due to the pervasive use of Internet in communication between diverse, identifiable and network connected components, Internet of Things (*IoT*) is the buzzword of choice for such an environment.

*IoT* [1] is a framework of connected computing devices such as electronic equipments, vehicles, machines and human assisted devices, operating and communicating with each other to offer a seamless set of services. In fact, the significant use of *IoT* could exponentially increase due to its application in the development of Smart Cities [2]. The technology could be prevalently used in the communication of various interacting devices of such cities. These devices need to provide important composed services and are required to inter-exchange stateful information. The devices could be uniquely identified in an Internet environment and they also use various communication technologies to share and interact with each other.

The field of Aspect-Oriented Software Development (*AOSD*) [3] is popularly known as Aspect-Oriented Programming (*AOP*). It allows the development of software to clearly and independently encapsulate the core and cross-cutting concerns as individual of functionality. The unique method of improvement in modularity was proposed by Gregor Kiczales [2] in the Palo Alto Research Center. Quite a number of researchers [5]–[7] have been working on inferring the quality of its application with a new set of metrics for measuring the core design properties of the software design. But, the acceptance of the methodology requires a thorough study of its application in various instances, where there is a need for improvement in modularity.

Any modern software inherently requires modules to connect itself with the network for inter-exchange of information. In such an environment, a Middleware becomes an essential component to manage a set of services. A Middleware was born when the common network services were refactored from the application logic and encapsulated as an independent set of services. These intermediate services are necessary to improve the modularity of the software and thereby increasing the quality of service in a networked environment.

In an *IoT* based environment, different network components such as sensors have to exchange their measured data with other components. Such exchange requires a communication component on both the ends with embedded set of important services. Some of these embedded services are security, handshaking, data transfer, etc. These services in the communication component are modelled as Middleware services and are vital in the successful functioning of the networked devices. Additionally, there are also services that are tangled and scattered across the above mentioned services of the Middleware.

As mentioned above, the Middleware of the *IoT* communication environment contains both core and cross-cutting functionalities. Hence, this is an ideal environment for refactoring these different concerns into independent units of functionalities. Since *AOSD* provides constructs and methods for the clear separation of the two concerns, this methodology can be applied in refactoring the Middleware services. The refactored services can be further modelled into classes and aspects of *AOP*. The most prevalently used and supported *AspectJ* programming language [8] was considered for the implementation of refactored concerns in a simulated environment of the Middleware services.

Cohesion is one of the important software design property reflecting in the attainment of modularity level for a software. Considering this property, it is always preferred to have a higher cohesion value for its components. In this work, the design property of cohesion is measured for the different design components and compared with equivalent *Java* and *AspectJ* versions of the Middleware software. A similar implementation and measurement has not been attempted by others. This work will signify the application of the methodology in the architecture of *IoT*. Based on the comparison, it was found that the cohesion of the classes have improved in the *AspectJ* version compared to the equivalent *Java* version.

The rest of the paper is organized as follows: Section II expands the existing work in the literature about the Middleware services and application of *AOP* in *IoT*. Section III expands the primary objectives of this research. The concept of modularity of Middleware in *IoT* is explained in Section IV. Section V explains the proposed measurement model for this research. The new set of proposed metrics for this research work is defined in Section VI. The metrics values obtained by applying them on to the simulated environment is tabulated and explained in Section VII. Section VIII discusses the inferences on applying the proposed metrics. Finally, Section IX concludes and provides pointers for the future scope of this research work.

## II. EXISTING WORK

Applying the concepts of *AOP* is an important area that requires a considerable amount of investigation. This will help in understanding the overall effect of applying the technique to the mainstream programming. Even though a number of authors have attempted to measure the effect of aspectizing a *non-AO* based software, work done on applying this unique technique on the *IoT* platform is very few in the literature.

In a work done by Maingret et. al., [8], event based handlers have been used to interconnect the dynamic external behaviors with the decision process of an *IoT* environment. In such a kind of environment, the dynamic external behaviors can be modeled as an *aspect* of *AOP* thereby decoupling them from the decision making process. Such a solution was not attempted by the authors to improve the overall modularity of the framework.

Razzaque et. al., [9] have done a survey on the work done by researchers on the Middleware of *IoT*. A number of authors specified in the paper have reported their findings on identifying the methods and technologies in the environment. But, there is no work done on improving the modularity of the Middleware software by incorporating techniques to neatly encapsulate the different concerns.

In a work done by Cerny [10], the author has suggested a typical roadmap in using *AOP* to effectively address with the cross-cutting concerns in the design of applications and integration of systems for the *SOA* and *IoT* environments. But no implementation and measurement attempts have been made to quantitatively assess the effect of such application in the Middleware of the *IoT* environment.

Wang et. al., [11] have used *Spring AOP* framework for the application of *IoT* and fog computing in community safety. According to the author this was done in a 5G environment, but the Middleware was not considered for the application of *AOP*. The framework used is similar to the *AspectJ* programming language but does not include the constructs in the level of the programming language. Also no quantitative measurements have been attempted to infer on the effect of software quality.

## III. FOCUS OF THE RESEARCH PROBLEM

Aspect-Oriented Software Development had focused on improving the modularity of software for the clear and neat encapsulation of core and crossing concerns as independent units of functionalities. But, the application of this methodology requires various implementations and measurements across different domain of software. Implementation of software using this methodology and development of metrics to quantify the modular constructs is the first step in evaluating this unique methodology.

In this paper, the objective is to remodel the software that has been designed using object oriented methodology into a higher modular Aspect-Oriented software. Since, the focus is on the Middleware of *IoT*, the constructs defined in the *Java version* of the *IoT* Middleware have been refactored and remodeled into its equivalent *AspectJ version*. New set of metrics have been defined to measure the effect of aspectization on the core design property of cohesion for the respective version. The design property of cohesion has been chosen since it is one of the most important property affecting the modularity of software.

## IV. MODULARITY OF MIDDLEWARE IN IoT

Two communicating entities in an *IoT* environment require a Middleware for the communication and exchange of acquired data through its sensors. Hence, a Middleware can be considered as the backbone for the design and development of *IoT* infrastructure. Fig. 1 depicts the typical role played by the Middleware in such an infrastructure.

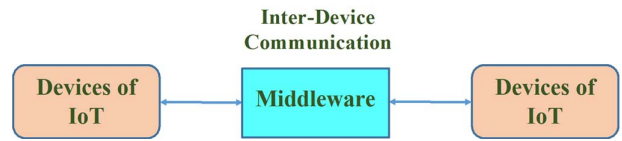

Fig. 1. Role of Middleware in IoT

The Middleware being an intermediary between two IoT devices consists of different types of functionalities embedded inside its framework. These functionalities provide necessary and vital services to the *IoT* infrastructure. A few of its cohesive set of functionalities are handshaking, data transfer and security. These are the core set of functionalities built as part of the Middleware services. There are also other set of functionalities that constitute the Middleware services defined in its framework.

The functionalities built as the Middleware services of *IoT* framework are independently designed and encapsulated into software modules. Such an independent modularization is necessary to improve the cohesiveness of the modules in the software. In spite of doing so, there are still set of functionalities that cut across the functionalities embedded in these modules. These functionalities are the scattered and tangled set of concerns that cannot be neatly encapsulated into independent units and thereby improving the modularity of the Middleware software. Hence, there is a need to introduce a software development methodology to enable the modularization of cross-cutting concerns into independent units of functionality.

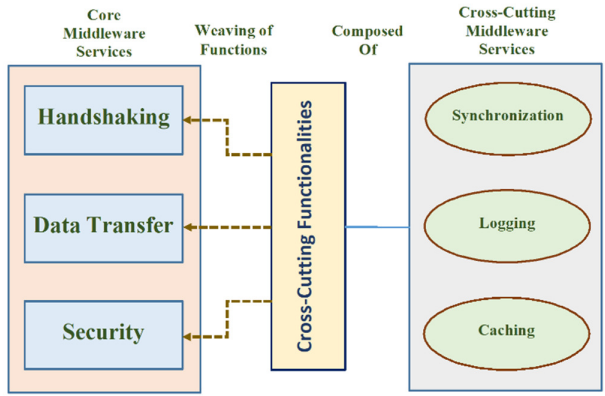

Fig. 2. Framework of *AO* in *IoT*

The significance in the need for modularization can be clearly visualized from the block diagram shown in Fig. 2. The core and the cross-cutting services are shown as separate units inside the rectangular boxes on both ends of the diagram. The core services considered are handshaking, data transfer and security functionalities. These are the important set of core services considered in this study, since they are the necessary and primary services of the *IoT* Middleware.

Similarly, the services shown in oval boxes are the composition of cross-cutting Middleware services. These tangled and scattered services are woven on the core Middleware services shown as rectangular boxes. The cross-cutting services considered are the modularized functionalities of synchronization, logging and caching. Again, these are the most important set of functionalities modeled as cross-cutting Middleware services.

## V. PROPOSED MEASUREMENT MODEL AND METRICS

Understanding the changes in software modeling needs the definition of a systematic and predictable quantitative measurement model. The measurement needs to be defined in such a way that it can accurately capture the software design properties in focus. The current study is about the effect of changing the design in the modelling of Middleware software. Hence, the proposed change in modelling of concerns using *AOP* needs to independently capture both the constructs that define the core and cross-cutting concerns.

The two different cohesion measurement models to measure the Object Oriented (*OO*) and Aspect Oriented (*AO*) versions of the *IoT* Middleware software is shown in Fig 3. The *Java* version of the software will map the functionalities using classes. Whereas the equivalent *AspectJ* version is able to independently model the core and cross-cutting concerns into separate set of functionalities. As specified in Fig. 2 the core functionalities considered in the design are handshaking, data transfer and security. In the *AO* version the functionalities of synchronization, logging and caching are considered and modeled as aspect in *AspectJ* language.

It is well known that any study on the effect of using a different software design model cannot be measured directly. Hence, there is a need for a well-defined and comprehensive quality model for its evaluation. The quality model can include the measurement of design property from which the effect on the quality can be inferred in a systematic manner. The design property of cohesion is considered in this study to measure and infer on the improvement in quality. The measurement models shown Fig. 3 considers cohesion for the measurement.

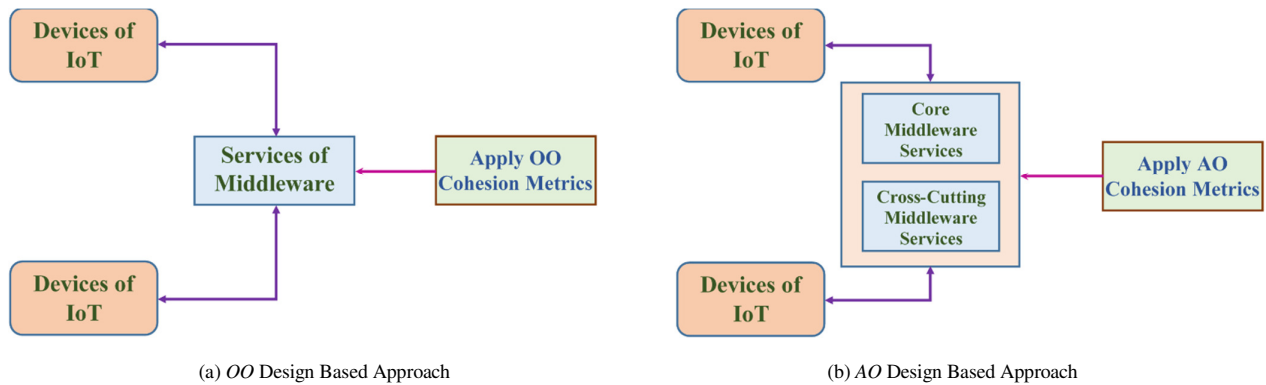

(a) *OO* Design Based Approach  (b) *AO* Design Based Approach

Fig. 3. Cohesion Measurement Model in Middleware Services of *IoT*

## VI. MEASUREMENT OF COHESION

Cohesion is one of the most important design property to be considered during the design of software. It is a core software design property that quantifies the ability of a software module to focus only on a single functionality. The spectrum for the cohesion metric value ranges from low to high or from 0 to 1. A software module that exhibits higher cohesion will have a value closer to 1 and if it exhibits lower cohesion, then the value will be close to 0. For example, if a software module is defined with two embedded functionalities, then the value of its cohesion is 0.5. Similarly, if three functionalities are embedded then its cohesion value is 0.33.

In the literature, various metrics have been proposed by researchers [12]-[15] in order to quantify cohesion of the given *AO* software design. These metrics were able to consider the different constructs defined in the *AO* design and implementation. For example, in *AspectJ* programming the *classes* and *aspects* are the two encapsulated modular entities considered for the measurement of cohesion. These constructs have to be individually measured to find the net cohesion value of an *AspecJ version* of software. The equivalent *Java version* will only contain the *classes* as the encapsulated entities and will be counted during the measurement.

The metrics considered for the measurement of *Cohesion Index* (*CoI*) in this study is shown in Equations 1 and 2.

Consider the following units before specifying the metrics:

Let *p* and *q* be the number of *classes* in the *Java* and *AspectJ versions* of *IoT* Middleware and *f(i)* and *f(j)* are the number of functionalities defined in class *i* and class *j* of *Java* and *AspectJ versions* respectively.

Let *r* be the number of *aspects* in the *AspectJ version* of *IoT* Middleware and *f(k)* be the number of functionalities in aspect *k* of the *AspectJ* version.

*Cohesion Index (Java classes)*,

$$CoI(J) = \frac{\sum_{i=1|j=1}^{p|q} 1/f(i|j)}{p|q} \qquad (1)$$

*Cohesion Index (AspectJ aspects)*,

$$CoI(AJ) = \frac{\sum_{k=1}^{r} 1/f(k)}{r} \qquad (2)$$

The metric *CoI(J)* for the *Java classes* and *CoI(AJ)* for the *AspectJ aspect* of the *IoT* Middleware have been normalized and can range between the values 0 to 1. This method enables to compare the two equivalent, but different model based versions to be compared to evaluate the effect of aspectizing an *OO* based software.

VII. MEASURED VALUES OF PROPOSED METRICS

In this paper, a new set of metrics have been proposed to measure the cohesion of *Java* and *AspectJ versions* of a software. Further, the software that simulates the Middleware of the *IoT* environment has been developed using *Java* and *AspectJ* programming languages. The functionalities embedded in the versions have been explained in Section II of this paper.

The first metric, cohesion of the *Java classes CoI(J)*, is applied on both the *Java* and *AspectJ versions* of the *IoT* Middleware software. Cohesion of the *AspectJ aspects*, *CoI(AJ)* metric, has been applied only on the *AspectJ version* of the *IoT* Middleware software. The respective measured values of the two proposed metrics, *CoI(J)* and *CoI(AJ)* are tabulated in Table I.

TABLE I. MEASURED METRIC VALUES FOR JAVA AND ASPECTJ VERSIONS OF IoT MIDDLEWARE

| S. No. | Version | CoI(J) | CoI(AJ) | Average of CoI(J) & CoI(AJ) |
|---|---|---|---|---|
| 1 | IoT Middleware Java version | 0.19 | - | 0.19 |
| 2 | IoT Middleware AspectJ version | 0.38 | 1 | 0.69 |

During the measurement, the first metric, *CoI(J)* has been applied on both the versions of the *IoT* Middleware because, the *AspectJ version* consists of both *classes* and *aspects*. The *classes* were used to encapsulate the core concerns and the *aspects* were used to encapsulate the cross-cutting concerns of the software. The vice versa is not possible as the *Java version* modularizes the core and cross-cutting concerns in a tangled and scattered form of software code.

VIII. DISCUSSION ON MEASUREMENTS

Understanding the effect of aspectizing an *OO* based *IoT* Middleware software requires careful definition and application of measurements. As discussed in the previous sections, a new set of metrics have been defined to the measure the core design property of cohesion for the *Java* and *AspectJ* versions of the *IoT* Middleware. These metrics have been applied to the respective versions and the value are tabulated in Table I. In order to visually understand the effect of aspectization the measured values of the proposed metrics have been statistically represented using a bar graph shown in Fig. 4.

The following are the set of inferences derived from this quantitative assessment which provides insights on the effect of aspectizing *IoT* Middleware towards the higher level software quality attributes:

- Value of the *CoI(J)* metric has increased in the *AspectJ version* of the Middleware software compared to the *Java version*. This implies that the modularity of functional components modeled as *classes* in the *AspectJ version* is better than the *Java version* of the Middleware. The reason for such a change in the values is because the tangled and scattered functionalities modelled in the classes of the *Java version* have now been refactored into more modular *aspects* and woven at the respective join points in the *classes*.

- Notably, *CoI(AJ)* of the *AspectJ version* has shown a high cohesion with the maximum value of 1. This has happened because the tangled and scattered concerns present in the *Java version* of the Middleware is now refactored as independent units of functionality. In our implementation, the ability of the *AspectJ* implementation to independently encapsulate the functionalities of synchronization, logging and caching has helped to aspects to attain the highest values of cohesion metric. This value in turn also suggests that aspects are components with high degree of modularity and reusability.

- The average value of the *CoI(J)* and *CoI(AJ)*, has naturally increased in the *AspectJ version* of the Middleware compared to its equivalent *Java version*. This increase reflects the ability of the *AOP* technique to improve the overall cohesive strength of the modules modelled with the required functionalities of the Middleware. Such a kind of modeling will help in the overall quality of the software thereby, decreasing the maintainability of the modules encapsulated as *classes* and *aspects*.

To summarize, aspectization of the Java version of the *IoT* Middleware which resulted into the equivalent *ApsectJ version* has shown measured sign of increase in reusability, maintainability and modularity, which are some of the core higher level quality attributes of a software.

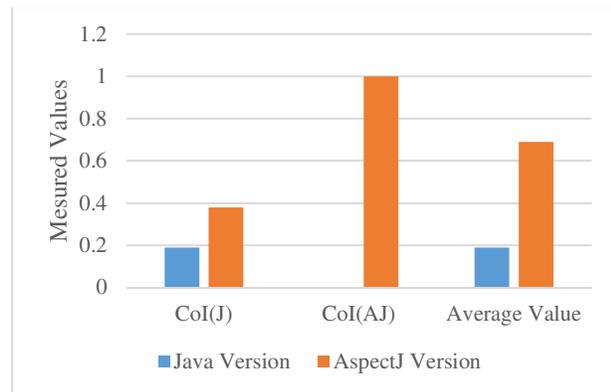

Fig. 4. Metric Values for *Java* and *AspectJ* versions of *IoT* Middleware

## IX. CONCLUSION AND FUTURE WORK

The powerful and flexible *AOSD* methodology, which was introduced in the late 90s, has sustained and grown in spite of other introducing other similar methodologies. The proponents of *AOSD* have done a considerable amount of research work to evaluate and learn about the positive effects and difficulties in adopting this methodology. In the last 20 years, a large number of researchers have attempted to apply and quantitatively evaluate using new set of metrics. Due to this a significant amount of research content is available in the literature related to aspectization.

Based on the need to quantitatively assess the effects of aspectizing a *non-AO* (*OO*) software, this research has focused on applying the technique in the infrastructure of *IoT*. In particular, the Middleware of the *IoT* infrastructure has been considered for the application of aspectization. The core and cross-cutting concerns of the functionalities embedded in the Middleware have been considered the application. Both *Java* and *AspectJ* versions of the *IoT* Middleware were developed using a simple simulation environment.

Further, to evaluate the effect of aspectizing an *OO* based software, the design property of cohesion has been considered for the measurement. A new set of metrics have been defined to measure the effect of aspectizing an *OO* based Middleware software. The first metric was used to measure the functionalities embedded in the constructs in the *Java* classes and the second metric was used to measure the *cross-cutting* functionalities defined in the *aspects* of *AOP*. The first metric, *CoI(J)* was applied to both the *Java* and *AspectJ versions* of the *IoT* Middleware, whereas the second metric, *CoI(AJ)* was applied only to the *AspectJ* version of the Middleware.

Based on the measurement it was found that the modularity and reusability of the functionalities has improved in the *AspectJ version* of *IoT* Middleware due to the increase in the cohesion values. This work can be further extended by increasing the number of functionalities considered in the *IoT* Middleware for aspectization. Further, the core software defined in the *IoT* devices of the infrastructure can also be considered understand the overall effect of applying *AOSD* methodology.